\documentclass[journal]{vgtc}                     

\usepackage{stfloats}
\usepackage{wrapfig}
\usepackage{xspace,xpunctuate}
\usepackage{amsmath}
\newcommand{\ie}{{i.e.,}\xspace}
\newcommand{\eg}{{e.g.,}\xspace}

\newcommand{\etc}{{etc\xperiod}\xspace}

\onlineid{0}

\vgtccategory{Research}

\vgtcpapertype{Application}

\title{AttentionViz: A Global View of Transformer Attention}

\author{%
  Catherine Yeh, 
  Yida Chen,
  Aoyu Wu,
  Cynthia Chen,
  Fernanda Vi\'egas, and 
  Martin Wattenberg
}

\authorfooter{
  \item Authors are with Harvard University. Vi\'egas  and Wattenberg are also with Google, but this work was done at Harvard. Emails: \{catherineyeh, yidachen, aoyuwu, fernanda, wattenberg\}@g.harvard.edu, cynthiachen@college.harvard.edu
}

\abstract{Transformer models are revolutionizing machine learning, but their inner workings remain mysterious. In this work, we present a new visualization technique designed to help researchers understand the \textit{self-attention} mechanism in transformers that allows these models to learn rich, contextual relationships between elements of a sequence.
The main idea behind our method is to visualize a joint embedding of the \textit{query} and \textit{key} vectors used by transformer models to compute attention. Unlike previous attention visualization techniques, our approach enables the analysis of global patterns across multiple input sequences. We create an interactive visualization tool, AttentionViz (demo: \url{http://attentionviz.com}), based on these joint query-key embeddings, and use it to study attention mechanisms in both language and vision transformers. We demonstrate the utility of our approach in improving model understanding and offering new insights about query-key interactions through several application scenarios and expert feedback. 
}


\teaser{
  \centering
  \includegraphics[width=\linewidth]{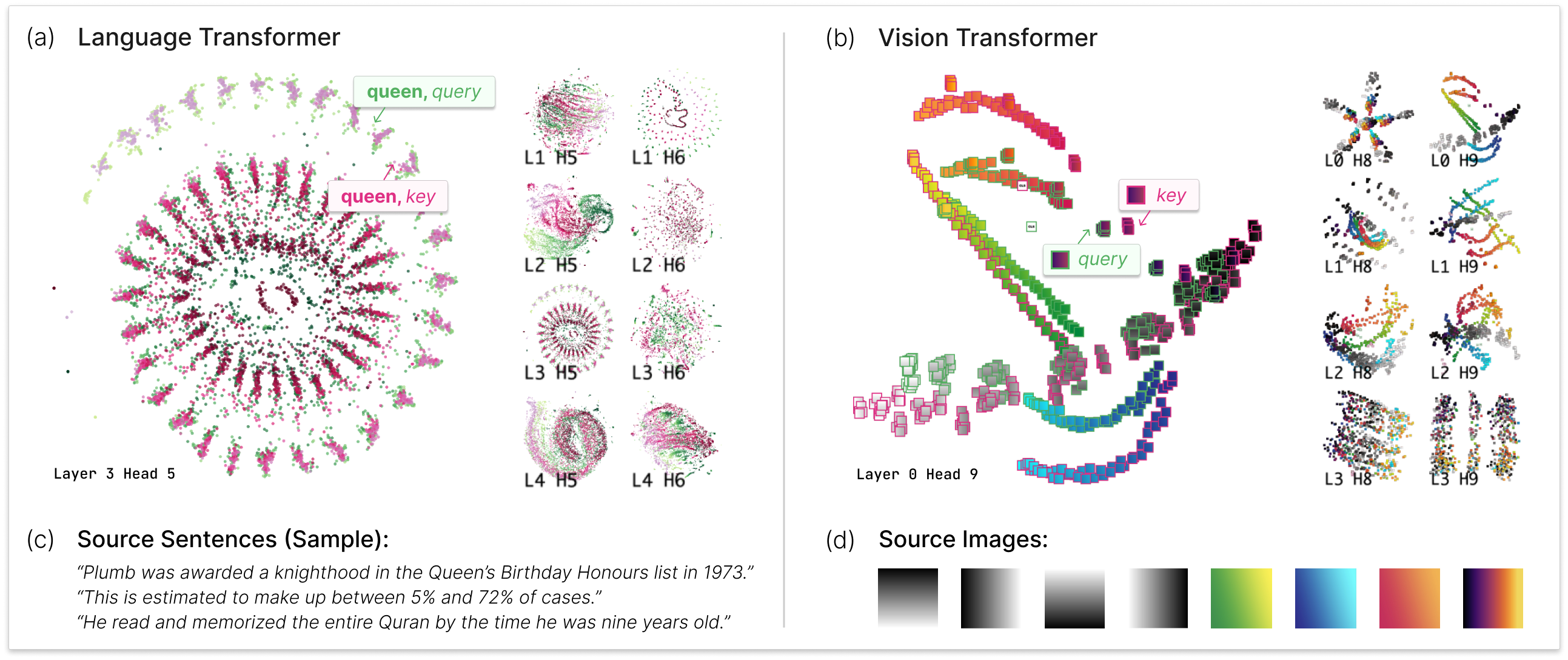}
    \caption{
  AttentionViz, our interactive visualization tool, allows users to explore transformer self-attention at scale by creating a joint embedding space for \textcolor{ForestGreen}{\textit{queries}} and \textcolor{magenta}{\textit{keys}}. \textbf{(a)} In language transformers, these visualizations reveal striking visual traces that can be linked to attention patterns. Each point in the scatterplot represents the query or key version of a word, as denoted by point color. Users can explore individual attention heads (left) or zoom out for a ``global'' view of attention (right). \textbf{(b)} Our visualizations also divulge interesting insights in vision transformers, such as attention heads that group image patches by hue and brightness. 
  Border color denotes query embeddings of a patch (\textcolor{ForestGreen}{green}) or key embeddings (\textcolor{magenta}{pink}).
  \textbf{(c)} Sample input sentences (from~\cite{jiang2020neural}) and \textbf{(d)} images (synthetic dataset) are provided for reference.
  }
  \label{fig:teaser}
}

\graphicspath{{figs/}{figures/}{pictures/}{images/}{./}} 

\usepackage{tabu}                      
\usepackage{booktabs}                  
\usepackage{lipsum}                    
\usepackage{mwe}                       

\usepackage{float}

\usepackage{mathptmx}                  

\begin{document}


\firstsection{Introduction}

\maketitle

The transformer neural network architecture~\cite{vaswani2017attention} is having a major impact on fields ranging from natural language processing (NLP)~\cite{devlin2018bert,radford2019language} to computer vision~\cite{dosovitskiy2020image}. Indeed, transformers are now deployed in large, real-world systems used by hundreds of millions of people (\eg \href{https://stability.ai/}{Stable Diffusion}, \href{https://openai.com/blog/chatgpt}{ChatGPT}, \href{https://blogs.microsoft.com/blog/2023/03/16/introducing-microsoft-365-copilot-your-copilot-for-work/}{Microsoft Copilot}). However, the mechanisms behind this success remain somewhat mysterious, especially as new capabilities continue to emerge with increasing model complexities and sizes~\cite{dehghani2023scaling,wei2022emergent}. A deeper understanding of transformer models could help us build more reliable systems, troubleshoot problems, and suggest avenues for improvement.

In this work, we describe a new visualization technique aimed at better comprehending how transformers operate. (We include a brief introduction to transformers in Sec.~\ref{sec:background}.) The target of our analysis is the characteristic transformer \textit{self-attention} mechanism, which allows these models to learn and use a rich set of relationships between input elements. Although attention patterns have been intensively studied, previous techniques generally visualize information related to just a single input sequence (\eg one sentence or image) at a time. Typical approaches create bipartite graph~\cite{vig2019bertviz,vaswani2018tensor2tensor} or heatmap ~\cite{liu2018visual,hoover2019exbert} representations of attention weights for a given input sequence.

Our method offers a higher-level perspective, in which we can view the self-attention patterns of many input sequences at once. One inspiration for this approach is the success of tools such as the Activation Atlas~\cite{carter2019activation}, which allows a researcher to ``zoom out'' to see an overview of a neural network, then drill down for details. In our case, we seek to build a kind of ``attention atlas'' that can provide researchers with a rich and detailed view of how a transformer's various attention heads operate. The primary new technique is visualizing a joint embedding of the \textit{query} and \textit{key} vectors used by transformers, which creates a visual signature for an individual attention head.

To illustrate our technique, we implement AttentionViz, an interactive visualization tool that allows users to explore attention in both language and vision transformers. 
AttentionViz affords exploration through multiple levels of detail (Fig.~\ref{fig:teaser}), providing both a global view to see all attention heads at once and the ability to zoom in on details in a single attention head or input sequence.

We demonstrate the utility of our technique through several application scenarios with AttentionViz and  domain expert interviews. For concreteness, we focus on what the visualization can reveal about a few widely-used transformers: BERT~\cite{devlin2018bert}, GPT-2~\cite{radford2018improving}, and ViT~\cite{dosovitskiy2020image}. We find several identifiable ``visual traces'' linked to attention patterns in BERT, detect novel hue/frequency behavior in ViT's visual attention mechanism, and uncover potentially anomalous behavior in GPT-2. User feedback also supports the wider applicability of our approach in visualizing other embeddings at scale.

To summarize, the contributions of this work include:
    \begin{itemize}
        \item A visualization technique for exploring attention trends in transformer models based on joint query-key embeddings.
        \item AttentionViz, an interactive tool that applies our technique for studying self-attention in vision and language transformers at multiple scales.
        \item Application scenarios and expert feedback showing how AttentionViz can reveal insights about transformer attention patterns.
    \end{itemize}
\section{Background on Transformer Models}\label{sec:background}

The \textit{transformer}, introduced in~\cite{vaswani2017attention}, is a neural network architecture designed to operate on sequential input.
A full description of transformers is beyond the scope of this paper, but a few concepts are critical for understanding our work. 
First, a transformer receives as input a set of vectors (often called \textbf{embeddings}). Embeddings can represent a variety of input types. In text-based transformers, they correspond to words or pieces of words; in vision transformers, they encode patches of pixels.

The network iteratively transforms these vectors via a series of \textbf{attention layers}, each of which 
moves information between pairs of embeddings. The name ``attention'' suggests that not all embeddings will be equally related; certain pairs will interact more strongly--\ie pay more ``attention'' to each other. Attention layers determine which pairs should interact, and what information should flow between them.

For example, in a transformer operating on the words of the sentence, ``The brown capybara is sleeping now,'' one might expect high attention (and information flow) between embeddings for ``capybara'' and ``is,'' but not between ``brown'' and ``now.'' The \textit{self-attention} mechanism, which is our focus in this paper, allows transformers to learn and use a rich set of relationships between elements of a sequence, yielding significant performance improvements across various NLP and computer vision tasks~\cite{devlin2018bert,radford2018improving,dosovitskiy2020image}.

There may be different reasons for embedding pairs to attend to each other. For instance, in our example sentence,``brown'' and ``capybara'' are linked by an adjective-noun relation, while ``capybara'' and ``is'' form a subject-verb relation. To allow for several relation types, transformer attention layers consist of multiple \textbf{attention heads}, each of which can represent a different pattern of attention and information flow. 

Each attention head computes its own attention pattern using a bilinear form computed from a query weight matrix $W_Q$ and key weight matrix $W_K$. Concretely, for two embedding vectors $x$ and $y$,
attention is related to the scaled inner product of a \textbf{query vector}, $W_Qx$, and a \textbf{key vector}, $W_Ky$. Letting $d$ be the dimension of $W_Ky$, we have:

$$f(x,y) = \frac{1}{\sqrt{d}}\langle W_Qx, W_Ky \rangle $$

Given embedding vectors $\{x_1, x_2, \ldots, x_n\}$, we compute the
self-attention between $x_i$ and the other vectors using the softmax function:
$$ attn(x_i, x_j) = \mathrm{softmax}_j(f(x_i, x), \ldots, f(x_i, x_n)) = e^{f(x_i, x_j)} / \sum_k  e^{f(x_i, x_k)} $$
Critically, this formula shows 
that the greater the dot product between the query and key vectors, the higher the final attention value will be, a fact that we rely upon in our joint embedding visualization.

There is much more to the transformer architecture than we cover here. In particular, 
we have only described the attention weighting between pairs of embeddings, and not the specific information that flows between them. (As discussed later, this is an area ripe for further investigation.) 
One last technical point is worth mentioning, however, since it will help interpret images later in the paper. 
The initial embeddings given to a transformer typically incorporate a vector representation of their ordering (for a 1D sequence) or spatial configuration (for a grid, as in vision transformers). For sequences, these position vectors are defined using trigonometric functions, and are located on a helix-like curve in high-dimensional space (see \cite{vaswani2017attention}). 

\subsection{Models Studied in this Paper}
We study three transformer models: BERT (language), GPT-2 (language), and ViT (vision). Each has been an important object of study in the machine learning community, and the three span a range of transformer architectures and applications.
\textbf{BERT}, or Bidirectional Encoder Representations from Transformers~\cite{devlin2018bert}, is a multi-layer transformer encoder. 
As a bidirectional model, BERT can attend to tokens (\ie input elements) in either direction.
\textbf{GPT-2}, or Generative Pre-trained Transformer 2~\cite{radford2019language}, is 
a multi-layer transformer decoder.
GPT-2 is a unidirectional model, meaning it only attends to previous tokens. 
\textbf{ViT}, or Vision Transformer~\cite{dosovitskiy2020image}, 
employs a self-attention-based transformer architecture 
by splitting images into ``patches'' and treating them like tokens in a sentence. Similar to BERT, ViT is a multi-layer, bidirectional transformer encoder. 
In this work, we look at ViT performance on 16x16 (\textbf{ViT-16}) and 32x32 (\textbf{ViT-32}) patch sizes.

\section{Related Work}
Many researchers have attempted to investigate the inner workings of transformers.~\cite{chi2020finding, miaschi2020linguistic} seek to understand the performance improvement from transformer-based language models by exploring learned linguistic representations, and~\cite{tenney2019bert} observed that BERT recapitulates classic steps in natural language analysis, from part-of-speech tagging to relation classification.
Another popular approach is mechanistic interpretability, \ie reverse engineering transformer models (\eg~\cite{elhage2021mathematical,olsson2022context,elhage2022toy}).
Attention, the backbone of transformers, has also been studied intensively. For example, attention appears to relate to syntactic structures in NLP systems~\cite{clark2019does, wang2022interpretability} and gestalt-like grouping in vision transformers~\cite{mehrani2023self}. Researchers have compared ViT's visual attention mechanism with convolutional filters as well, finding that attention is more robust against image occlusion, corruption, and high-frequency noise~\cite{naseer2021intriguing, park2022vision}. In our discussion of related work, we focus on visual approaches for studying transformer attention.

\subsection{Visualizing Attention in a Single Input Sequence}
Attention patterns naturally lend themselves to visualization, 
in both language and vision transformers
~\cite{park2019sanvis,derose2020attention,jaunet2021visqa,ma2022visualizing,caron2021emerging}.
These visualizations are largely focused on visualizing attention weights between query and key tokens \textit{in a single input sequence} using bipartite graphs (\eg~\cite{vig2019bertviz,vaswani2018tensor2tensor,liu2018visual,strobelt2018s}) or heatmaps (\eg
~\cite{ji2021usevis,liu2018visual,hoover2019exbert,reif2019visualizing,kovaleva2019revealing,aflalo2022vl,cordonnier2019relationship}). 

A few visualizations have been proposed that allow comparison across multiple models or layers. For instance, Attention Flows~\cite{derose2020attention} supports users in comparing attention within and across layers of BERT, as well as among attention heads given a single sentence. Dodrio~\cite{wang2021dodrio} uses a grid view, applied to single inputs, that enables direct comparison of attention heads. Another system,
VisQA~\cite{jaunet2021visqa}, visualizes attention at different heads for visual question-answering tasks by showing heatmaps of language self-attention, vision self-attention, and language-vision cross-attention. Even in these model-comparison systems, however, an analyst must look at different inputs, \textit{one at a time}, to identify and verify patterns for a given attention head.

\subsection{Beyond Single Inputs: Visualizing Embeddings and Activation Maximization}

It is natural to seek patterns that hold across multiple inputs. One technique that has proved effective toward this goal is visualizing collections of embedding vectors from multiple input sequences
~\cite{hohman2018visual,smilkov2016embedding,wang2018comparison,boggust2022embedding, hohman2019s, sivaraman2022emblaze}. For example, \cite{reif2019visualizing} visualized BERT embeddings for the same word used in many different contexts, and found clusters that corresponded to word senses. In an exploration of syntax processing, \cite{chi2020finding} visualized embeddings from a multilingual BERT model and once again found meaningful clusters that helped with interpretation. LMFingerprints ~\cite{sevastjanova2022visual} uses a tree-based radial layout to compare embedding vectors across different language models.

A second technique, used for vision transformers in \cite{ghiasi2022vision,yang2021transpose}, aims to find images which maximize activations of particular units. Applied to embedding vectors, this technique produces clearly interpretable results. The authors note, however, that when applied to query and key vectors the technique does not seem to produce useful results.

\subsection{Gaps in the Literature}
We note three gaps in the existing literature which motivate our work.    

First, visualizing embedding vectors has been shown to be an effective technique for analyzing patterns across multiple inputs, but we know of no systematic attempt to visualize query and key embeddings in transformer models.~\cite{chefer2021transformer} also argues that the intermediate artifacts of self-attention, such as queries and keys, are underexplored. These observations motivate our joint query-key embedding technique.

Second, although visualization techniques have been proposed to compare multiple embeddings (\eg~\cite{li2018embeddingvis,arendt2020parallel,boggust2022embedding}), these methods are often limited to a few embeddings and cannot address our needs of comparing embeddings at different transformer heads and layers. 
Thus, we design a global matrix view to visualize query-key embeddings at scale.

Finally, bipartite graph representations have proven helpful in analyzing NLP-based transformers, but we have not seen them applied to vision tasks. We explore this direction by creating bipartite-style visualizations to study image attention patterns in ViT.
\section{Goals \& Tasks}\label{sec:goals_tasks}
The overarching aim of this work is to design a novel visualization technique that allows exploration of global attention trends in transformer models.
To collect some initial feedback about this idea and learn more about user needs, we talked with 5 machine learning (ML) researchers (4 Ph.D. students and 1 professor) interested in model interpretability. During these individual interviews, we asked experts to describe their current practices and challenges when working with transformers and how attention visualizations could aid in their research objectives. We will refer to these experts as \textbf{E1-5}.

\subsection{Goals}
Ultimately, our conversations with experts yielded three main goals:

\textbf{G1 - Understand how self-attention informs model behavior.} 
Overall, all 5 experts wanted to better understand the behavior of different attention heads and what transformer models are learning through their characteristic self-attention mechanism.
Thus, they expressed the desire to be able to quickly and easily explore attention patterns.
\textbf{E2} explained that ``attention is still pretty closed-box and there's lots of mysteries,'' so gaining a deeper understanding of transformer attention patterns could provide insights into ``why large language models fail at reasoning tasks and math,'' for example. 

\textbf{G2 - Compare \& contrast attention heads.} \textbf{E5} mentioned that visualizing differences in attention heads could help with hypothesis generation, which is the first step in their research process: ``Visualization can help formulate hypotheses to test and get an intuitive sense of what transformers are doing.'' Additionally, 3 experts (\textbf{E1, E2, E5}) noted that attention head comparison would be useful for model pruning and editing purposes. That is, if two attention heads appear to behave similarly, perhaps one could be removed without significantly impacting model performance. In the words of \textbf{E1}, comparing heads might allow us to ``find parts of the model that are actually useful.''

\textbf{G3 - Identify attention anomalies.} Four researchers (\textbf{E2-5}) wanted to identify irregularities and potential behavioral issues with transformers through attention pattern exploration. This information could then be used for model debugging purposes. 
For instance, 
\textbf{E4} said ``visualizing attention could help you notice when the model is looking at the wrong thing, even if the result is correct.'' \textbf{E3} agreed, reiterating the importance of debugging especially in the context of model training: ``Training often fails and dies, but it's hard to understand why it fails or produces unexpected behavior.''

\subsection{Tasks}\label{sec:tasks}
Given these goals, we developed the following set of design tasks:

\textbf{T1 - Visualize attention heads at scale.} To help users quickly explore model behavior \textbf{[G1]} and easily compare \& contrast attention patterns \textbf{[G2]}, our tool simultaneously visualizes self-attention heads across transformer layers. 

\textbf{T2 - Explore query-key interactions.} \textbf{E1} and \textbf{E4} expressed the desire to better understand query-key pairing information toward improving their understanding of transformer self-attention. Thus, our tool further supports attention pattern comparison \textbf{[G2]} and anomaly detection \textbf{[G3]} through visualizing query-key interactions.

\textbf{T3 - Probe attention at multiple levels.} Our tool allows for local and global comparisons of attention \textbf{[G2]} by providing visualizations at the \textit{sentence/image}, \textit{head}, and \textit{model} levels. The flexibility of switching between multiple views in a single interface also facilitates knowledge discovery \textbf{[G1]} and helps users identify model irregularities \textbf{[G3]}.

\textbf{T4 - Customize model and data inputs.} AttentionViz is easily extendable to new transformers and datasets, affording quick visual comparison \textbf{[G2]} and synthesis of attention patterns \textbf{[G1]} across different models and modalities (language \& vision).
\section{Query/Key Embeddings \& Design of AttentionViz}
To address these goals and tasks, we build a tool called AttentionViz. The primary technique used by our tool is a visualization of the joint embedding of query and key vectors for each attention head. In this section, we first describe the motivation and mathematics that underlie this technique, then discuss the design of the full application.

\begin{figure}
    \centering
    \includegraphics[width=0.9\linewidth]{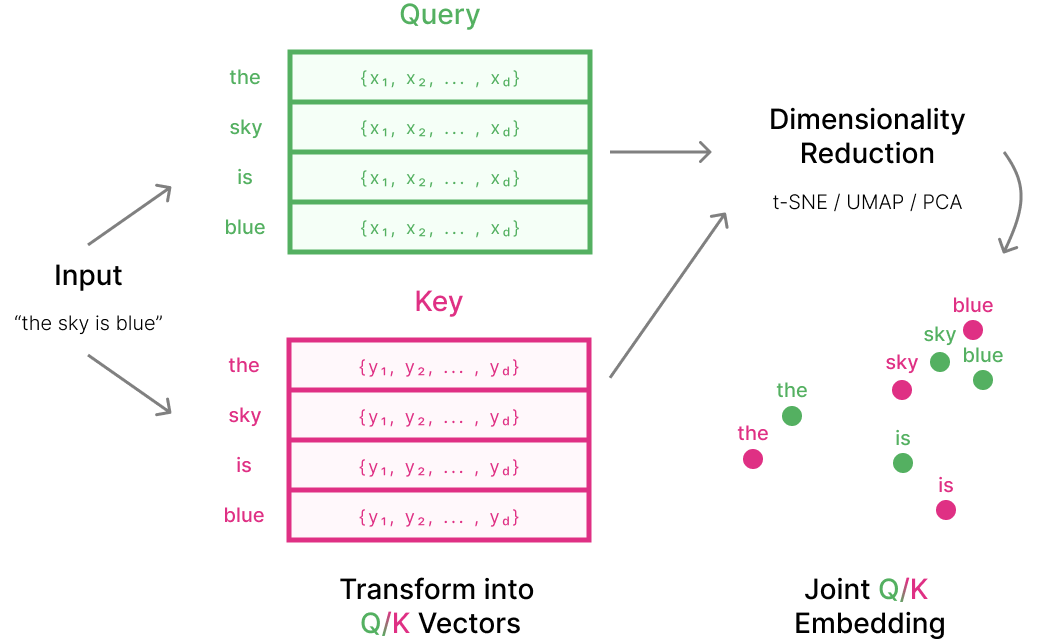}
    \caption{Creating a joint query-key embedding space for a single attention head. In the NLP case, given an input sentence, we first transform each token into its corresponding query and key vector. Then, we use t-SNE/UMAP/PCA to project these $1 \times d$ vectors into 2D/3D scatterplot coordinates. For the BERT, GPT-2, and ViT models used, $d = 64$.}
    \label{fig:joint_embedding}
\end{figure}
\subsection{Visualizing Query/Key Embeddings}\label{sec:joint_embeddings}

The technique behind AttentionViz is relatively simple, although as we describe below, it requires two mathematical tricks to be effective. Recall that each transformer attention head transforms input embeddings into query vectors and key vectors by applying matrices $W_Q$ and $W_K$, respectively (Sec.~\ref{sec:background}). These matrices project the original vector embeddings to a lower dimensional space, essentially selecting a particular type of information from the higher-dimensional vector embeddings. Therefore, by inspecting the query and key vectors, one might hope to learn what information is selected by $W_Q$ and $W_K$.

A central observation is that the relative positions of query and key vectors can offer clues about how attention will be distributed, since attention coefficients depend on the dot product between queries and keys. 
To see why, consider a hypothetical situation where query and key vectors always have the same norm. Then, closer distances would directly relate to higher attention coefficients. In practice, query and key vectors vary in norm, so the relationship between dot product and distance is not precise. However, as described in the following sections, we can arrange for this relation to be surprisingly close.

Fig.~\ref{fig:joint_embedding} illustrates our technique with a synthetic example of a single attention head in a language transformer. To create the joint embedding, we first obtain the query and key vector representation of each token in a given sentence (Sec.~\ref{sec:background}). Then, we use one of three dimensionality-reduction methods to project these high-dimensional vectors onto a shared, lower-dimensional subspace: \textbf{t-SNE}~\cite{van2008visualizing}, \textbf{UMAP}~\cite{mcinnes2018umap}, or \textbf{PCA}~\cite{Jolliffe1986}.
The output from these dimensionality-reduction algorithms is a 2D/3D scatterplot, where each point represents a single query or key token. The same process can be used to create joint embeddings for ViT attention heads, where each token is an image patch. 
By default, we visualize queries in \textcolor{ForestGreen}{\textit{green}} and keys in \textcolor{magenta}{\textit{pink}}. However, there are multiple color encodings users can choose from (see Sec.~\ref{sec:color}),
and other palettes can be easily substituted into the system.

Our joint embeddings allow users to explore the fine-grained interactions between queries and keys, and the shape of these plots can often serve as visual indicators of the underlying self-attention patterns (see Sec.~\ref{sec:eval}). Each dimensionality-reduction technique produces different patterns for a given dataset, offering unique insights and use cases.

\begin{figure}[t]
    \centering
    \includegraphics[width=0.85\linewidth]{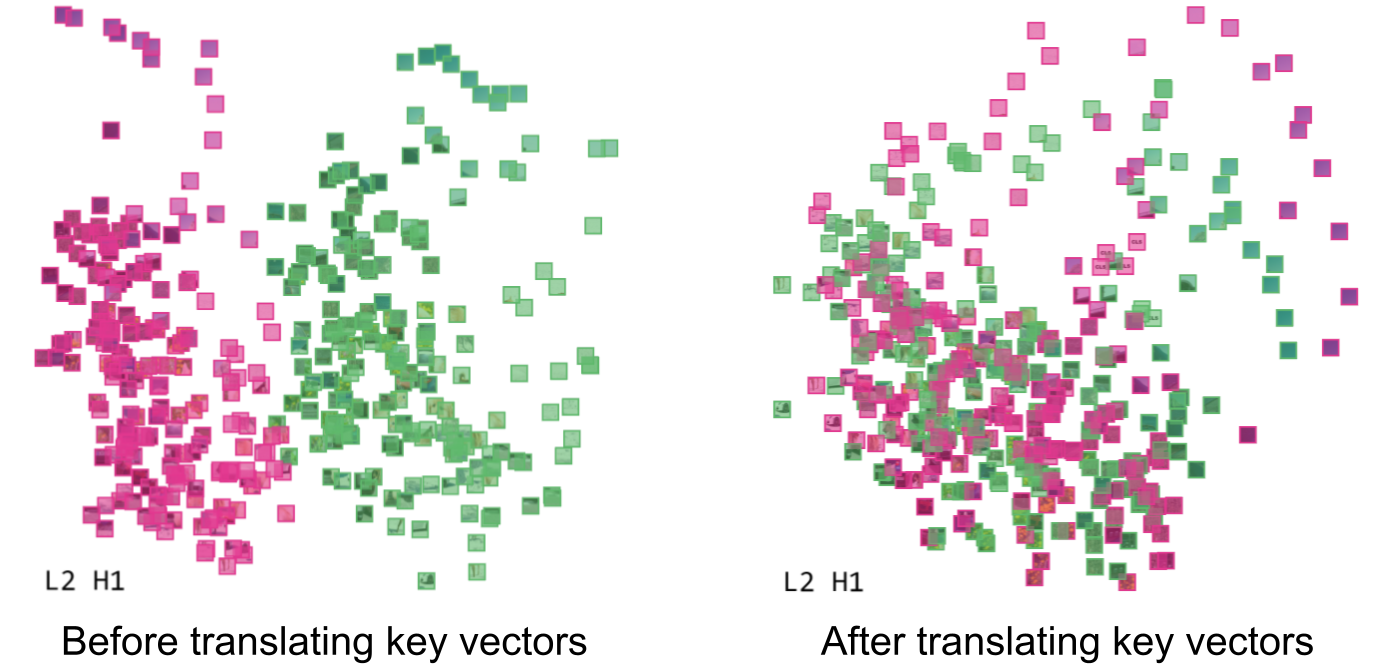}
    \caption{\textbf{Left:} original queries and keys in joint embedding space. \textbf{Right:} Increased overlap after translating keys to align query and key centroids.}
    \label{fig:translating-key-vectors}
\end{figure}
\subsubsection{Vector Normalization}\label{sec:key_norm}
While designing AttentionViz, we noticed two ``free parameters,'' which can be varied without losing any information. Tuning these parameters creates a closer relationship between embedding distance and attention weights, and greatly improves the readability of the visualization. These normalizations are applied prior to dimensionality reduction (Fig.~\ref{fig:joint_embedding}).

\textbf{Key Translation:} Query and key vectors are sometimes well separated in our visualizations (Fig.~\ref{fig:translating-key-vectors}, left).  
This separation makes it difficult to directly compare query and key embeddings. However, a simple mathematical trick allows us to move these embeddings closer together, without affecting the
self-attention computation for any given input sequence. In particular, note that the softmax function is \textit{translation invariant}: \ie for any constant $a$, we have $\mathrm{softmax}_j(x_1 + a, x_2 + a, \ldots) = \mathrm{softmax}_j(x_1, x_2, \ldots)$. Now, consider a query vector $x$ and key vectors $y_1, \ldots, y_n$. For any vector $v$, we have:
\begin{equation*}
\begin{split}
\mathrm{attention}_j(x) &= \mathrm{softmax}_j(\langle x, y_1 \rangle, \langle x, y_2 \rangle, \ldots) \\
&= \mathrm{softmax}_j(\langle x, y_1 \rangle + \langle x, v \rangle, \langle x, y_2 \rangle + \langle x, v \rangle, \ldots) \\
 &= \mathrm{softmax}_j(\langle x, y_1 +v \rangle, \langle x, y_2 +v \rangle , \ldots) 
\end{split}
\end{equation*}
where the second step follows by translation invariance. This implies that without changing attention patterns in any given input, we can translate all key vectors such that the query and key distributions for each attention head have identical centroids. This makes it much easier to compare queries and keys (Fig.~\ref{fig:translating-key-vectors}, right).

\textbf{Scaling Queries and Keys:} In some transformers, such as GPT-2, we observed cases where the average query norm was very different from the average key norm. This difference makes it hard to interpret key-query relations in a joint embedding. Mathematically, it indicates a poor relationship between dot product and distance; visually, it means queries might be a tiny cluster, surrounded by a loose cloud of keys.

\begin{figure}[t]
    \centering
    \includegraphics[width=0.9\linewidth]{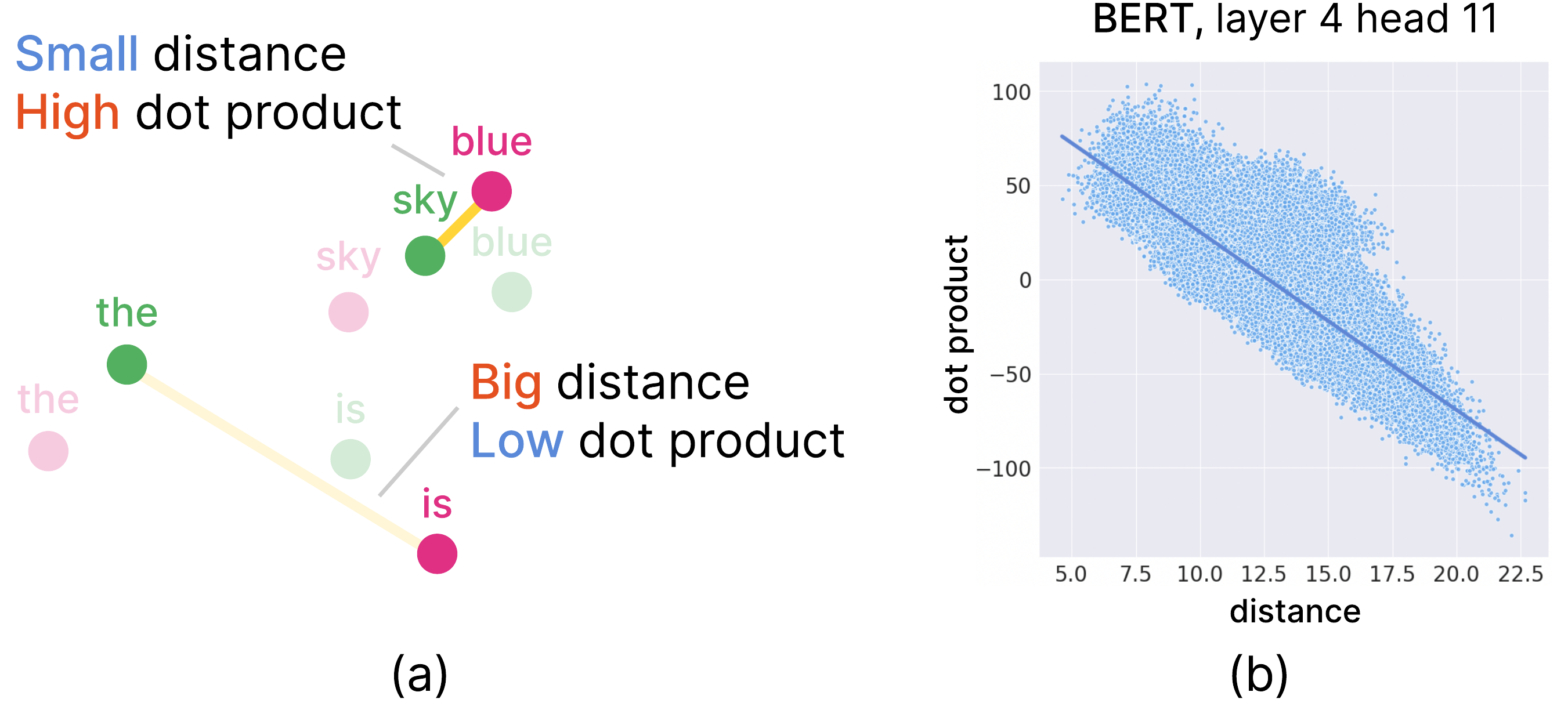}
    \caption{\textbf{(a)} Ideal distance-attention relationship, where query-key pairs with higher dot products are closer in the joint embedding space.
    \textbf{(b)} Example attention head with a strong, negative correlation (\textbf{-0.983}) between query-key distance and dot product in BERT.
    }
    \label{fig:correlation}
\end{figure}

Luckily, scale is another ``free parameter'' of the system.
Self-attention levels depend only on dot products of query and key vectors, so if we scale all query vectors by a factor of $c \neq 0$, and all key vectors by a factor of $c^{-1}$, the attention values are unchanged. This allows high-attention query-key pairs to be closer together in our joint visualizations as depicted in Fig.~\ref{fig:correlation}a. (A subtle point: on its own, scaling leaves cosine distance unchanged; however, in combination with translation normalization it has a nontrivial effect.) 

To determine the optimal value of $c$, we can define a \textit{weighted correlation} metric that places heavier weight on query-key pairs with smaller distances, since we care most about nearby queries and keys in the joint visualization. We can thus choose a scale factor $c$ such that the weighted correlation between query-key dot products and distances is maximized. This scaling method allows for the distances in the joint embedding space to most accurately represent the actual attention values between queries and keys.

\subsubsection{Distance as a Proxy for Attention}\label{sec:distance_attention}
As explained above, ideally, if a query-key pair has a large, positive dot product (corresponding to a high final
self-attention value), they should be placed closer together in the embedding space, and vice versa (Fig.~\ref{fig:correlation}a). Thus, we expect distance to be \textit{inversely correlated} with attention in our joint query-key embeddings.
We study this potential link by computing the Spearman rank correlation between cosine distance and dot product for each attention head in BERT, GPT-2, and ViT. We also experimented with using Euclidean distance as our distance metric when creating t-SNE and UMAP projections of queries and keys, but this generally led to weaker distance-dot product correlations.

Across multiple datasets and models, the relationship between distance and attention holds fairly well. For example, with Wiki-Auto data~\cite{jiang2020neural}, the mean correlation between query-key distances and dot products is \textbf{-0.938} for BERT and 
\textbf{-0.792} for GPT. An example result from BERT is shown in Fig.~\ref{fig:correlation}b.
On the set of COCO images used~\cite{lin2014microsoft}, the mean correlation is \textbf{-0.873} 
for ViT-32 and \textbf{-0.884} 
for ViT-16. 

\subsection{Color Encodings}\label{sec:color}
To visualize different properties of queries and keys, AttentionViz offers various color encodings. The default option colors points by \textit{token type}, \ie query or key.
For vision transformers, users can color by image patch \textit{row} or \textit{column} to visualize positional patterns (Fig.~\ref{fig:increasing-attention-distance}). Since images encode their own color information, we also allow users to view the \textit{original} patches without additional styling elements (Fig.~\ref{fig:machines-rods-cones}).

\begin{figure*}[t]
    \centering
    \includegraphics[width=\linewidth]{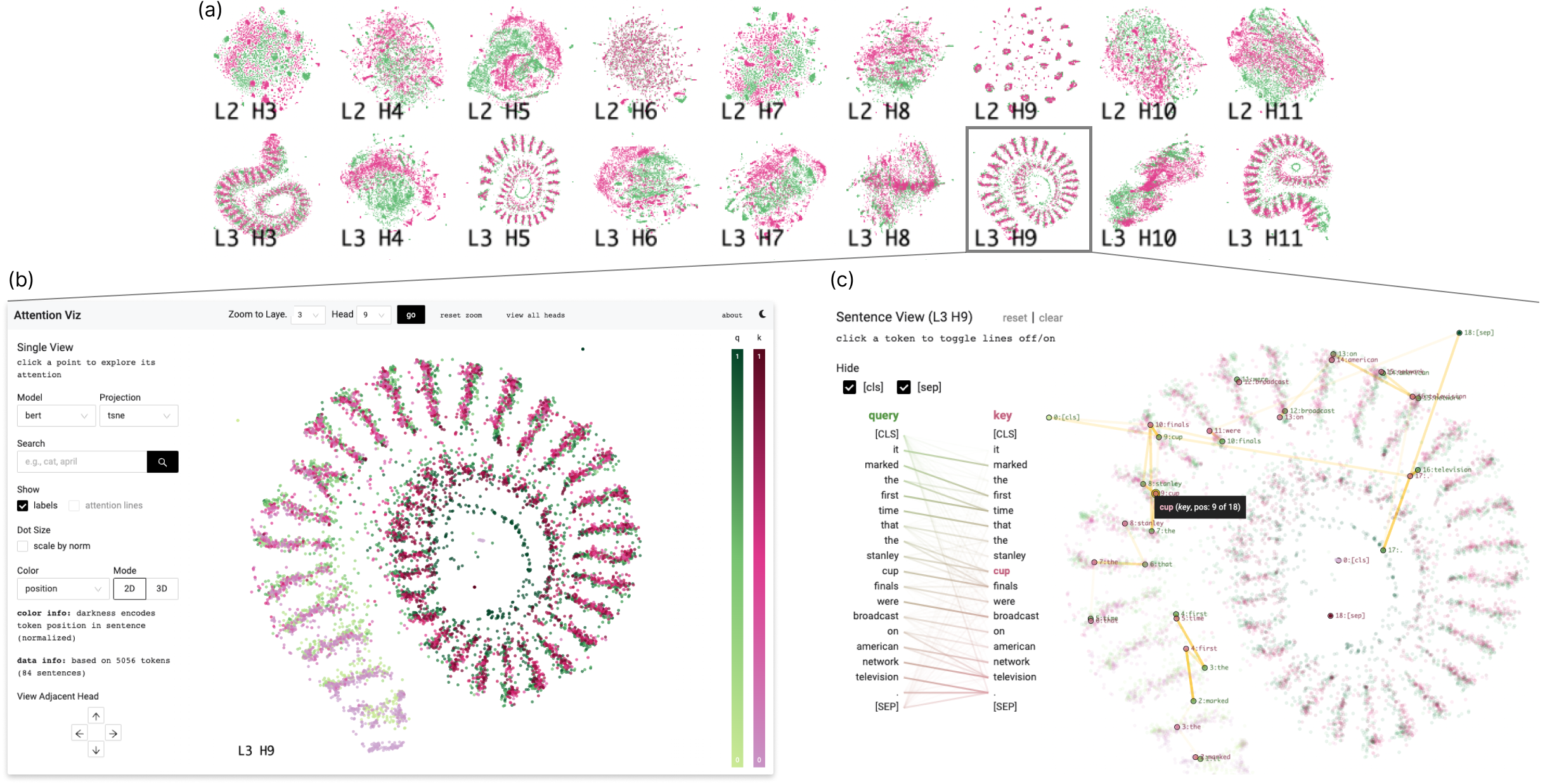}
    \caption{Connecting form to function in BERT. \textbf{(a)} In Matrix View, there are several spiral-shaped plots in layer 3. \textbf{(b)} By zooming into one such head (\textit{L3 H9}) using Single View, we can see positional attention patterns by using a light-to-dark color scheme that encodes position in the input sequence. \textbf{(c)} These patterns can be confirmed by exploring sentence-level visualizations.}
    \label{fig:spiral}
\end{figure*}

For language transformers, we support two positional color schemes: \textit{normalized} and \textit{discrete}. To compute normalized position, we divide each token's position in a sentence by the sentence length to produce a continuous color scale. Lighter hues denote tokens closer to the beginning of the sentence (Fig.~\ref{fig:spiral}b). Our discrete position encoding takes each token's position and applies the modulo operator to get its remainder when divided by 5. Thus, the $1^{st}$ and $6^{th}$ tokens receive the same color, the $2^{nd}$ and $7^{th}$ tokens receive the same color, \etc. 
We use the same five colors to encode queries and keys at different positions, using darker hues for the former.
Although this scheme introduces ambiguity for sentences with length > 5, we find our discrete coloring helpful in seeing relationships based on small offsets in position (e.g., queries paying attention to keys one step away as in Fig.~\ref{fig:traces}, left). Ambiguities are also easily resolved by hovering over tokens, and in our explorations, we did not see patterns with offsets greater than 2 or 3.
Users can color by \textit{query/key norm} as well (Fig.~\ref{fig:gpt_anom}a).

\subsection{Views}\label{sec:views}
AttentionViz provides three main interactive views for attention exploration: \textit{Matrix View}, \textit{Single View}, and \textit{Sentence/Image View}.  

\begin{figure}[t]
    \centering
    \includegraphics[width=\linewidth]{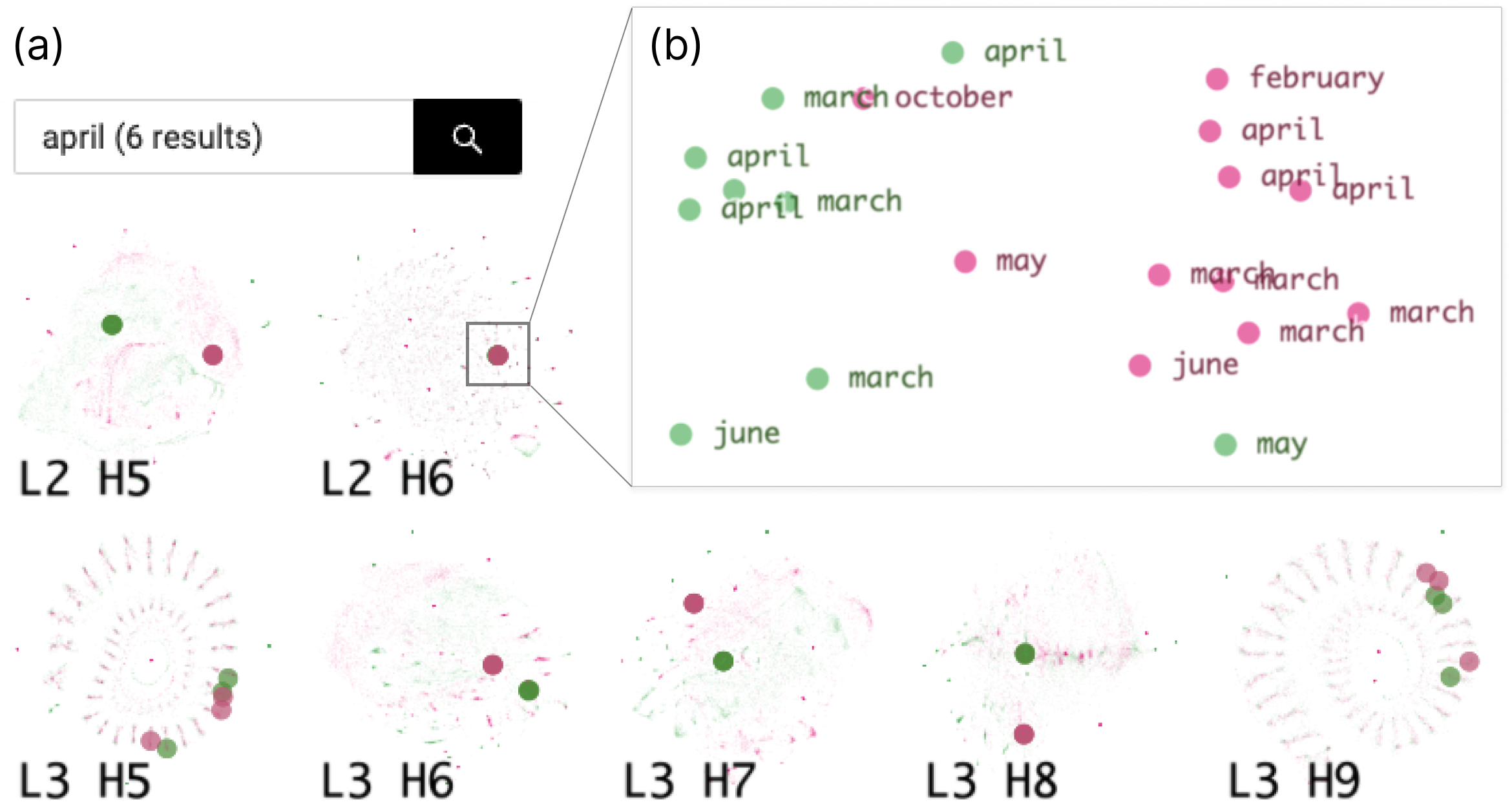}
    \caption{Exploring attention patterns with global search. \textbf{(a)} Heads with fewer clusters of search results often demonstrate more semantic behavior, while heads with dispersed results focus more on token position. \textbf{(b)} Zooming into L2 H6, a head with one main result cluster, we indeed see a large group of semantically related query and key tokens.}
    \label{fig:search}
\end{figure}
\subsubsection{Matrix View}
The initial view in AttentionViz is \textit{Matrix View}, which uses small multiples to visualize all the attention heads in a transformer at once (Fig.~\ref{fig:spiral}a), directly addressing \textbf{[T1]} and \textbf{[T3]}. Each row corresponds to a model layer, moving from earlier layers at the top of the interface to later layers at the bottom. With this ``global'' perspective, users can more easily scan for patterns across different transformer layers and heads, compared with single-plot (\eg~\cite{vig2019bertviz}) or instance-level visualizations (\eg~\cite{boggust2022embedding,smilkov2016embedding}). All the models used in this work had the same architecture: 12 layers x 12 heads per layer = 144 attention heads in total, but our system scales to other dimensions.

In Matrix View, users can view the joint query-key embeddings created with 
t-SNE, UMAP, or PCA. They can also switch between model types (\ie BERT, GPT-2, ViT-16/32) or datasets \textbf{[T4]}, explore different color schemes, and view the resultant plots in 2D or 3D. 
Matrix View supports a global search feature (Fig.~\ref{fig:search}a), which helps highlight patterns in token locations across different heads and offers another way to analyze attention at scale (see Sec.~\ref{sec:eval}).

\subsubsection{Single View}

Users can click on any plot in Matrix View to zoom into \textit{Single View} (Fig.~\ref{fig:spiral}b), which affords exploration of a single attention head in closer detail \textbf{[T3]}.
Like Matrix View, users can switch between colorings, dimensions, projection modes, datasets, and models in Single View \textbf{[T4]}; all graphical changes sync between views to facilitate comparison. The user can click on a point to highlight all tokens in the corresponding input sequence, spotlighting the relevant queries and keys in our joint embedding space. Users also have the option to project attention lines onto the scatterplots, which connect query and key tokens (Fig.~\ref{fig:spiral}c). We only show the top 2 attention weights for each token to enhance readability. Our attention lines feature supports \textbf{[T2]} and offers a new way to visualize attention patterns in transformers at the head level.

In Single View, users can also search for tokens and use our labelling feature to uncover semantic patterns in the data, similar to~\cite{smilkov2016embedding}. 
For instance, in Fig.~\ref{fig:search}b, search reveals that query/key tokens with similar meanings are placed together in the joint embedding for this BERT head, indicating strong attention between them (Sec.~\ref{sec:distance_attention}).

\begin{figure}[t]
    \centering
    \includegraphics[width=0.9\linewidth]{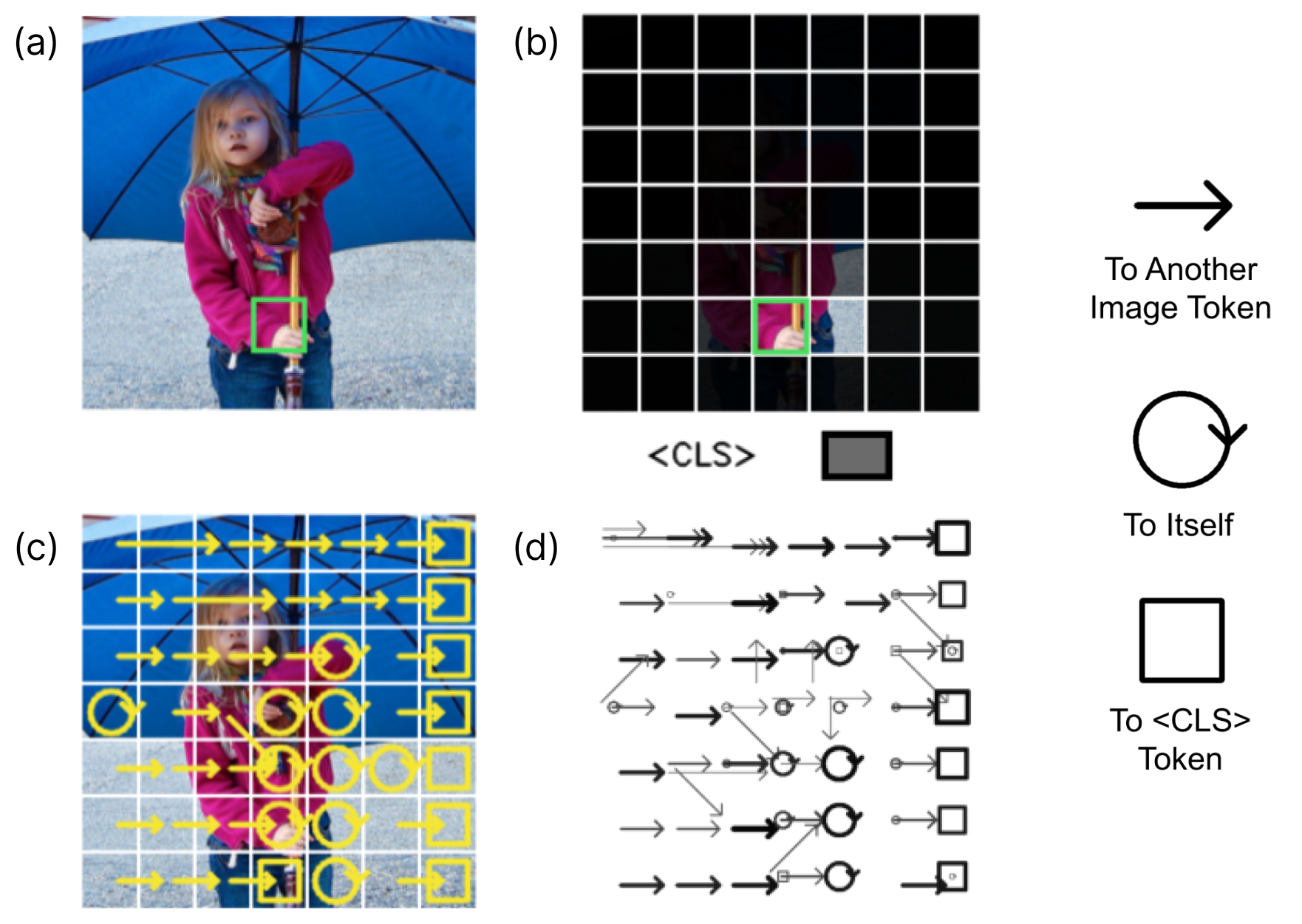}
    \caption{ViT Image View. \textbf{(a)} Original image. \textbf{(b)} Transparency attention heatmap. We highlight the selected token (query) with a green border. \textbf{(c)} Overlaid attention arrows. \textbf{(d)} Global attention flow. The square icon means an image patch has the strongest attention connection with the \texttt{<CLS>} token, which is not in the original image.}
    \label{fig:image-view}
\end{figure}
\subsubsection{Sentence/Image View} 
\textit{Sentence/Image View} allows for exploration of the fine-grained attention patterns within a single sentence or image \textbf{[T2, T3]}. Both views are synchronized with Single View, matching the attention lines  overlaid on each query/key scatterplot for a smooth user experience.

\textbf{Sentence View.} When using BERT or GPT-2, users can click on a point in Single View to open Sentence View in the left sidebar, which displays a BertViz-inspired visualization of sentence-level attention with the clicked token highlighted~\cite{vig2019bertviz} (Fig.~\ref{fig:spiral}c). We also considered using heatmap visualizations (\eg~\cite{park2019sanvis}), but it seemed that the bipartite graph approach would offer greater readability and ease of pattern exploration for longer sentences.
The opacity of the lines connecting query tokens in the left column and key tokens in the right column signifies their corresponding attention strength. Hovering on a token highlights token-specific attention lines. To reduce the noise from classification tokens and separators in BERT, or the first token in GPT-2 (Sec.~\ref{sec:eval}), users can hide the attention lines from these special tokens. Other query and key tokens can also be toggled on/off, and all attention lines will be re-normalized accordingly. 
Users have the option of viewing the aggregate attention pattern for each attention head as well, to offer another layer of comparison (Fig.~\ref{fig:traces}a).

\textbf{Image View.} For image-based input in ViT, when users click on an image patch, the side panel displays its corresponding original image and highlights the clicked token with a colored border (Fig.~\ref{fig:image-view}a). Users also see an image overlaid with an attention heatmap, where the transparency indicates the attention weight between the clicked image patch and other regions of the image (Fig.~\ref{fig:image-view}b). 

Beyond visualizing the attention of a single token, Image View
allows users to explore the overall attention pattern within an image by showing arrowed attention lines between different image patches. We provide users with two options when visualizing the attention arrows. The first option overlays arrows on the top of original image patches, with each arrow representing the strongest attention connection between a starting image patch and destination patch (Fig.~\ref{fig:image-view}c). This creates a simplified bipartite attention graph for users to characterize the most important patterns within a specific head. The second option shows all strong attention connections (\ie $attn(x_i, x_j) > 0.1$) beside the original image, offering a more comprehensive view of attention (Fig.~\ref{fig:image-view}d). In this visualization, both opacity and line thickness are used to encode the strength of attention connections.
We also tried visualizing all weights between queries and keys to more closely mirror~\cite{vig2019bertviz}, but this often produced overcrowded, inscrutable results. 
\section{System Implementation}
To process model inputs and compute attention information, we use the \href{https://huggingface.co/docs/transformers/index}{Hugging Face Transformers} library and PyTorch. We use pre-trained implementations of BERT, GPT-2 (small), and ViT-16/32 with model weights from Google and OpenAI.
 For each NLP dataset, we randomly sample 200 sentences ($\sim$10k tokens per attention head, including both queries and keys).
 Due to the increased computational size of image attention data, 
 we display 10 images per head (1000 tokens) for ViT-32 and 4 images per head (1576 tokens) for ViT-16.
 Larger datasets can be inputted into AttentionViz at the cost of increased system latency. Currently, it takes $\sim$6 seconds to load each NLP dataset, and $\sim$10 seconds to load ViT-16 data.
After extracting query and key vector embeddings for each attention head, we generate the corresponding 2D/3D t-SNE, UMAP, and PCA coordinates 
(Sec.~\ref{sec:joint_embeddings}). 
To produce semantic labels (\eg ``dog''
or ``background'') for image patches in ViT, we use the DeepLabv3 segmentation model~\cite{chen2017rethinking}. 
In total, it takes $\sim$3 hours to preprocess each BERT/GPT-2 dataset on a NVIDIA A100 GPU; for ViT-32, it takes $\sim$30 minutes.

Our final AttentionViz prototype consists of a Python/Flask backend that communicates with a frontend written in Vue and Typescript.
The demo system is available at: \url{http://attentionviz.com}. Due to the large size of the data and browser memory constraints, we load pre-computed attention/projection information
via JSON files through the backend. For ViT, the backend also performs image manipulation (\eg patch highlighting and transparency adjustments) to display in the frontend. 
We use \href{https://deck.gl/}{Deck.gl} to visualize the resultant query-key joint embeddings. AttentionViz is highly extensible and model-agnostic, allowing users to add new transformers and datasets to the system.

\begin{figure}[t]
    \centering
    \includegraphics[width=\linewidth]{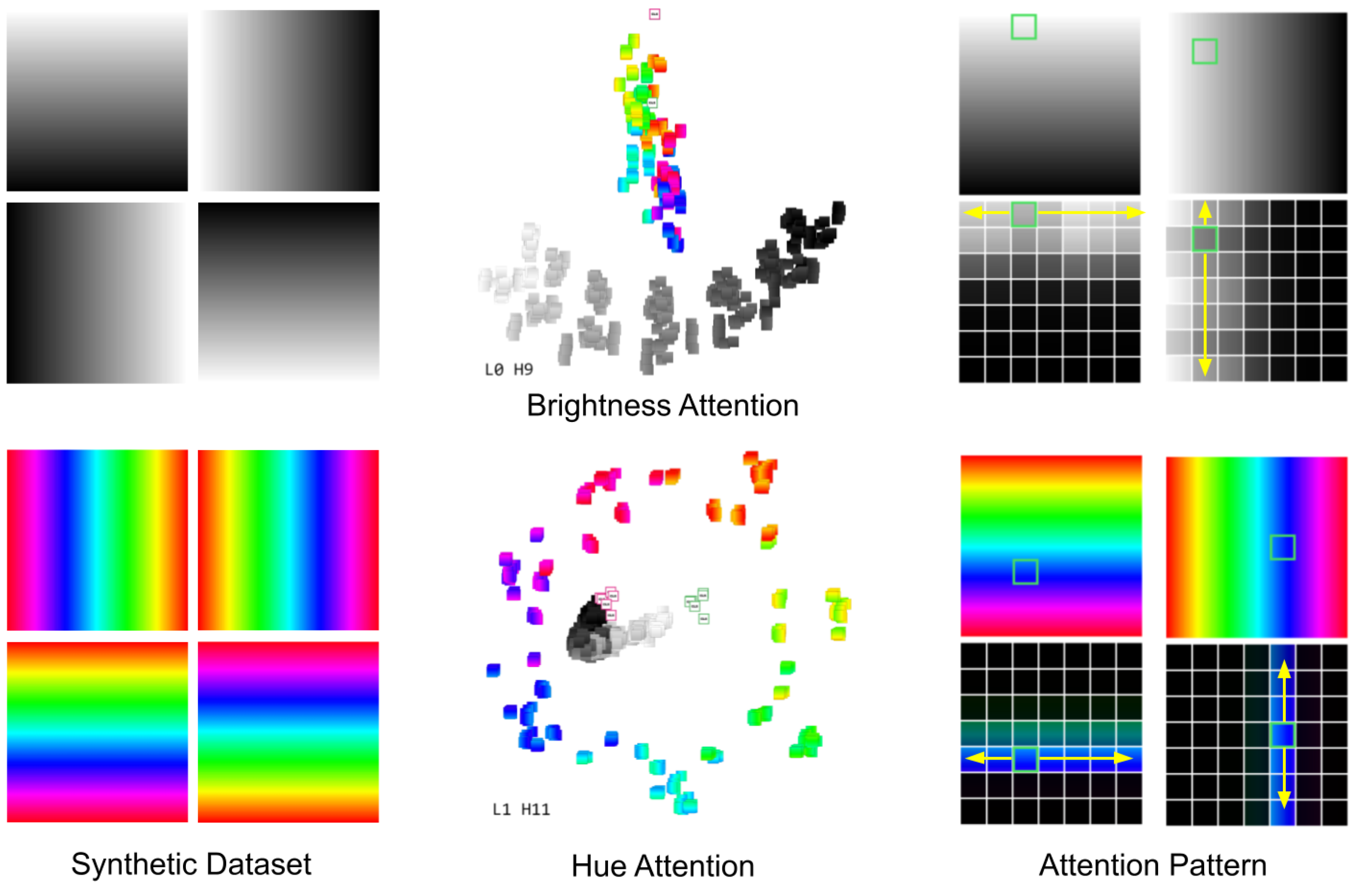}
    \caption{In ViT-32 Matrix View, we find two interesting visual attention heads: one head orders the black-and-white image tokens according to brightness, while the other aligns the colorful patches based on hue. Attention patterns shown in Image View confirm attention flow between patches with the same luminance.}
    \label{fig:machines-rods-cones}
\end{figure}
\section{Findings \& Evaluation}\label{sec:eval}
We illustrate the utility of AttentionViz with three application scenarios, as well as feedback from domain experts.
Our scenarios target the goals in Sec.~\ref{sec:goals_tasks}, and show how AttentionViz can offer insights about global self-attention trends in vision and language transformers.

\textbf{Data.} 
For BERT/GPT-2, we experimented with various NLP datasets but focus on two for our application scenarios. 
We use Wiki-Auto~\cite{jiang2020neural} as a baseline to sample general input sentences and SuperGLUE AX$_b$~\cite{wang2019superglue} to explore task-specific attention patterns for textual entailment.
For ViT, we sample images from
ImageNet~\cite{russakovsky2015imagenet} and Microsoft COCO~\cite{lin2014microsoft}, as well as synthetic image data.

\textbf{User Interviews.}
We invited \textbf{E2} and \textbf{E3} for a second round of interviews, and
included two new experts, \textbf{E6} (interpretability researcher) and \textbf{E7} (vision science Ph.D. student). As in Sec.~\ref{sec:goals_tasks}, all experts were interviewed individually.
We first gave experts a quick demo of our tool and shared some of our own findings, asking them to share any thoughts or insights (Sec.~\ref{sec:vis_att}-\ref{sec:anomalies}).  
Then, we asked for more general feedback about the main strengths, weaknesses, and novelties of AttentionViz (Sec.~\ref{sec:feedback}). We also asked experts about possible extensions or applications of this technique for visualizing embeddings at scale. 

\begin{figure*}[t]
    \centering
    \includegraphics[width=0.85\linewidth]{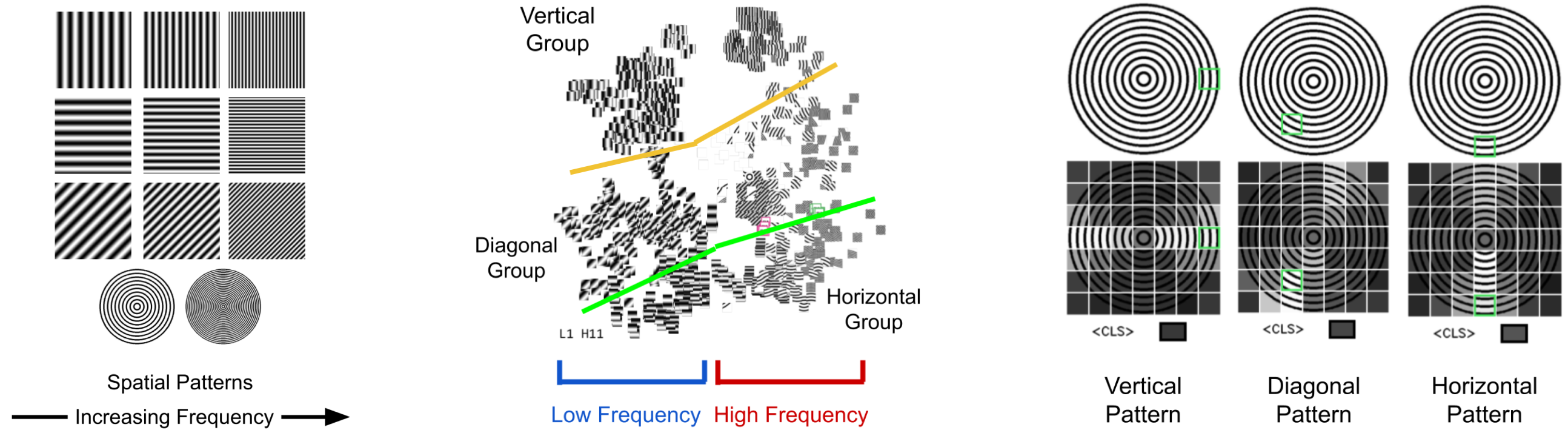}
    \caption{\textbf{Left:} Dataset of spatial patterns with different frequencies and angles. \textbf{Center:} In Single View, we observe that one attention head of ViT-32 arranges image tokens based on the frequencies and angles of their spatial pattern. \textbf{Right:} The attention heatmap in Image View further confirms these findings -- spatial patterns with similar angles pay greater attention to each other.}
    \label{fig:frequency-pattern}
\end{figure*}
\subsection{Goal: Understanding Machine Visual Attention}\label{sec:vis_att}
AttentionViz can be especially helpful in uncovering
 new insights about attention in vision transformers due to the inherently visual nature of image patch data \textbf{[G1]}.

\textbf{Hue/brightness specializations in visual attention.} 
We were curious if any visual attention heads specialize in either color-based and brightness-based patterns. To test this, we provided the pre-trained ViT-32 model with synthetic color and brightness gradient images (Fig.~\ref{fig:machines-rods-cones}), loading the resultant query and key tokens into AttentionViz. 

Browsing global PCA patterns in Matrix View, we identified two attention heads that resemble color and colorless vision. One head appears to align black-and-white image tokens based on brightness, and the other aligns colorful patches based on hue. 
Our dataset contains color and brightness gradient images in all orientations, and we see similar patches cluster together in the joint embedding space regardless of their position in the original images. The attention heatmap in Image View confirms these findings; tokens pay the most attention to other tokens with the same color or brightness. \textbf{E7} was intrigued by these results, having previously studied the color latent space of convolutional neural networks (CNNs), and expressed interest in using our tool to further explore the differences between CNN and ViT behavior.

\textbf{Frequency filtering and angle detection.} Frequencies and angles are low-level characteristics of image data. To investigate if the vision transformer has an attention head that associates visual patterns based on these features, we created images of sinusoidal signals with varying frequencies and orientations, processing them using our pretrained ViT-32 model. Examining the resultant query and key embeddings in Matrix View, we identified an attention head that separates image tokens based on their spatial pattern's frequencies (x-axis) and angles (y-axis) (Fig.~\ref{fig:frequency-pattern}). With Image View, we observed that tokens in the images of concentric circles are paying attention to other tokens with similar curvatures, further confirming that this attention head associates visual patterns based on their angles. 

\textbf{E7} said this result was interesting, but not too surprising given our hue/brightness findings, and was more curious about heads that do not exhibit this ``attend to similar patches'' behavior. One experiment they proposed was to study attention modulation, \eg if the same image patch (\eg vertical stripes) occurs in different contexts in two images (\eg zebra vs. umbrella), do we see unique attention patterns?

\begin{figure}[t]
    \centering
    \includegraphics[width=\linewidth]
    {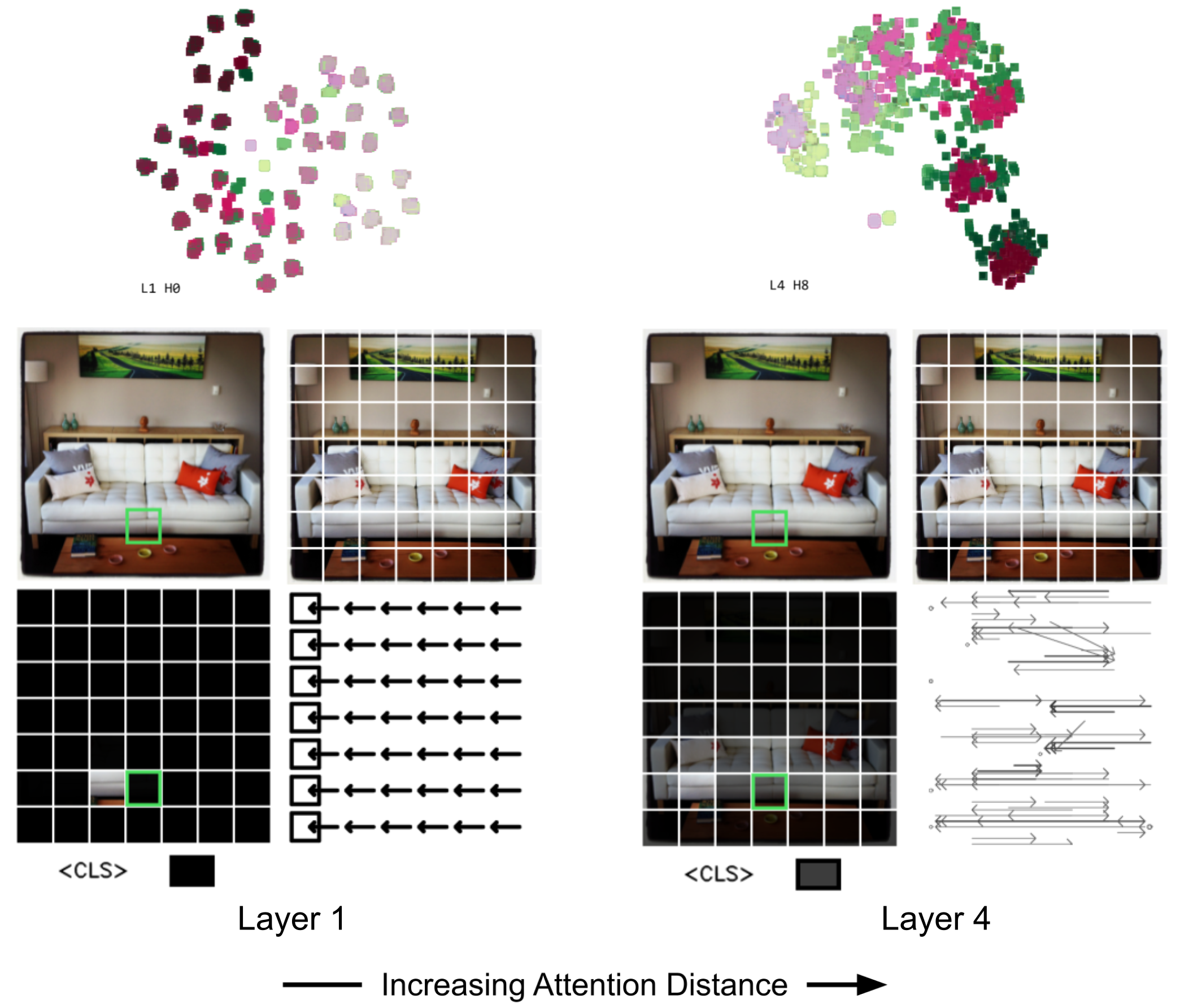}
    \caption{Coloring image patches by row highlights positional attention patterns in ViT-32. In \textbf{Layer 1}, tokens in the same row and adjacent columns form small clusters. Image View reveals a \textit{look at left} pattern. In \textbf{Layer 4}, large clusters of tokens form based on row positions. Using the arrowed lines, we see a wider, bidirectional attention flow.}
    \label{fig:increasing-attention-distance}
\end{figure}
\begin{figure*}[t]
    \centering
    \includegraphics[width=\linewidth]{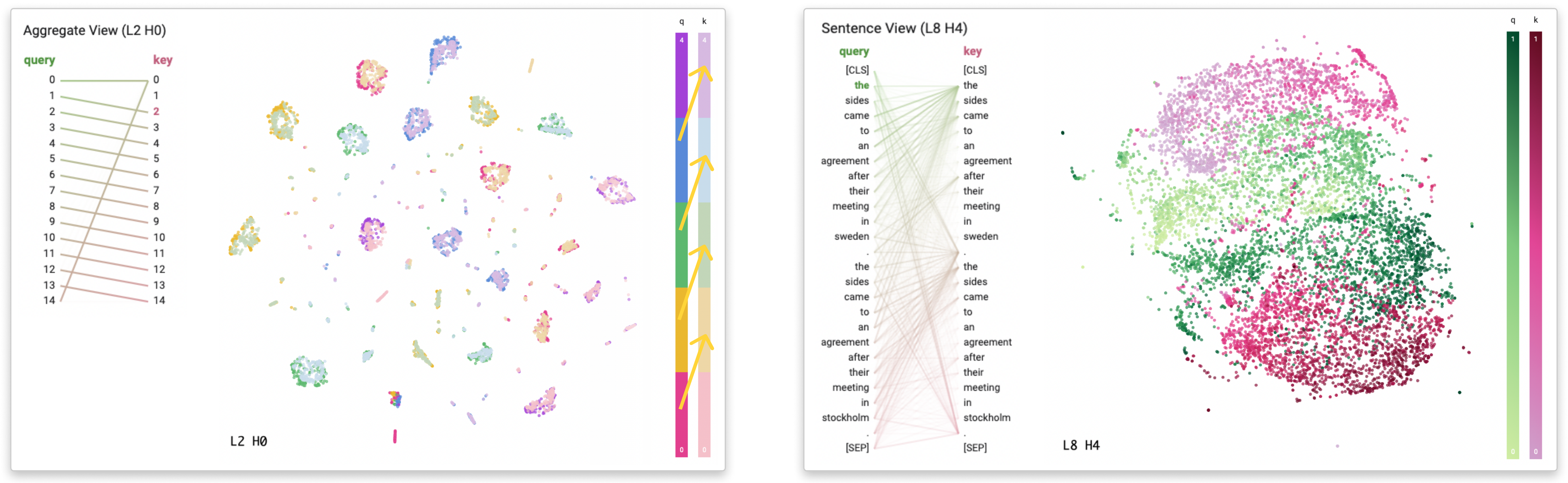}
    \caption{Other visual traces of attention.
    \textbf{Left:} Heads with small ``clumps'' often have even tighter positional patterns than spirals. Our discrete position encoding, which colors each token based on its position modulo 5, highlights a ``next-token'' attention trend.
    \textbf{Right:} Layered bands of queries and keys only appear with SuperGLUE AX$_b$ data~\cite{wang2019superglue}, indicating strong attention to text start, end, and midpoint.}
    \label{fig:traces}
\end{figure*}
\textbf{Increasing attention distance across model layers.} As noted in~\cite{dosovitskiy2020image}, self-attention attends more broadly across images in deeper layers of vision transformers. We confirmed this finding using our interactive, joint visualizations of query and key tokens in AttentionViz. With Matrix View, we colored patches by image ``row'' and ``column'' to find four attention heads in layers 1 and 2 of ViT-32 that group tokens with their nearest spatial neighbors: on their left, right, top, and bottom. In layers 3 and 4, we saw similar positional attention patterns, but image tokens pay attention to all the patches in the same row or column, beyond their nearest neighbors (Fig.~\ref{fig:increasing-attention-distance}). This suggests that unlike CNNs, which process images using a square filter, the self-attention mechanism in transformers often processes images row by row and column by column, analogous to an elongated filter.

\subsection{Goal: Finding Global Attention Traces}\label{sec:traces}
To understand how self-attention patterns vary across different heads in language transformers \textbf{[G2]}, we used AttentionViz to explore BERT.

\textbf{Positional attention signatures.} With TSNE, we observed 
several attention heads with unique shapes, \eg the spiral-shaped plots in layer 3 (Fig.~\ref{fig:spiral}a). For example, coloring layer 3 head 9 by \textit{normalized} position in Single View reveals that token position increases as we move from the outside to the inside of the spiral (Fig.~\ref{fig:spiral}b). We used Sentence View to examine this pattern more closely (Fig.~\ref{fig:spiral}c), confirming that there is a positional, ``next-token'' attention pattern. This ``spiral'' also reflects the initial ordering vector given to transformers (Sec.~\ref{sec:background}).

We then noticed other identifiable ``traces'' in Matrix View, finding that plots with small ``clumps'' also encode positional patterns (Fig.~\ref{fig:traces}, left), which we verified with our \textit{discrete} position coloring. 
The difference between ``spirals'' and ``clumps'' appears to be whether tokens attend selectively to others one position away, versus at several different possible positions (Fig.~\ref{fig:spiral}c). 
Similarly, we learned that in heads with high query-key overlap, tokens typically attend to themselves and other instances of the same token, exhibiting a ``look at self'' pattern.
Zooming into these heads, we see clear semantic clusters of nearby query-key pairs as shown in Fig.~\ref{fig:search}b, further supporting this observation. 

~\cite{lin2019open} shows that earlier transformer layers have the most information about linear word order, aligning with our findings and previous work such as~\cite{clark2019does, vig2019bertviz}. During our interviews, \textbf{E2}, \textbf{E6}, and \textbf{E7} immediately noticed these interesting geometries, particularly spirals, and were curious about how much of the observed structure is purely due to position. This inspired several follow-up experiment ideas from experts,
\eg manipulating or removing the positional embeddings in transformer models and seeing how our query-key visualizations change. 

\textbf{Task-specific traces.}
After visualizing multiple datasets with AttentionViz, we found that the shapes of joint embeddings are highly consistent across different NLP tasks. However, we did see one visual trace that only arises in some later layers of BERT with the SuperGLUE AX$_b$ data (Fig.~\ref{fig:traces}, right). Clicking on 
one such head (layer 8 head 9) and coloring by position, we observed a query-key ``sandwich,'' where keys and queries at the beginning of the text are stacked on top, followed by queries and keys at the end of the text in reverse order. 

Sentence View reveals that the start, middle, and end of the text receive the most attention. The overall plot shape and attention pattern suggests that these heads can identify a text's ``midpoint'' and differentiate between sentences, mirroring how in entailment tasks, two sentences are compared to see if they have similar meanings. Queries also mostly attend to keys in the same sentence.~\cite{kovaleva2019revealing,vig2019analyzing} shows how syntactic and task-specific information is most prominent in mid-to-later model layers, perhaps explaining the uniqueness of this trace.

\textbf{Global search patterns.} The aggregate search feature in Matrix View can also be used to quickly scan for and compare attention trends across heads \textbf{[G2]}. We found that patterns in the search results reflect the previously identified visual attention traces (Fig.~\ref{fig:search}a). For example, heads that are spiral-shaped or have small clumps of queries/keys have more dispersed search results, indicative of their underlying positional attention patterns. On the other hand, heads with the ``look at self'' attention pattern only have one cluster of search results, emphasizing the strong interaction between queries and keys of the same token. 

Even if a joint query-key embedding does not have a distinctive shape,
we see that
if there are only a few search result clusters, the head may display more \textit{semantic} behavior; otherwise, there is likely a \textit{positional} attention pattern.~\cite{tenney2019bert} notes that semantic information is spread across BERT's layers, which we confirmed with AttentionViz. All of our experts were particularly excited by this feature of our tool and its ability to facilitate attention pattern comparisons.

\subsection{Goal: Identifying Anomalies and Unexpected Behavior}\label{sec:anomalies}
Through interacting with the joint query-key embeddings in AttentionViz, we discovered some irregular model behaviors \textbf{[G3]}.

\begin{figure}[t]
    \centering
    \includegraphics[width=\linewidth]{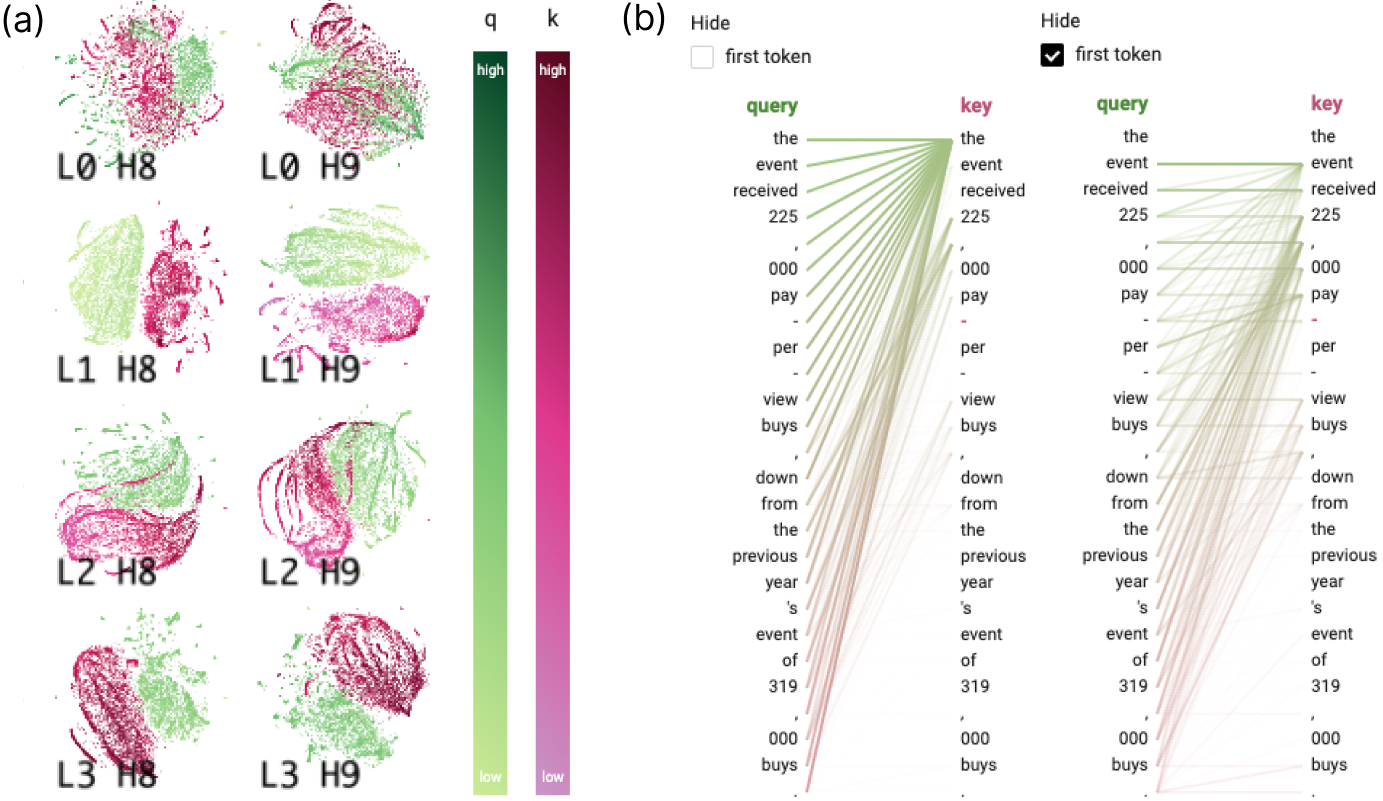}
    \caption{Anomalies in GPT-2. \textbf{(a)} In early model layers, we witness a significant disparity between query-key norms for many attention heads (e.g., \textit{L1 H8} prior to norm scaling). \textbf{(b)} Example of the prevalent ``attend to first'' pattern in later layers. Sentence View reveals latent attention behavior after hiding the first token.}
    \label{fig:gpt_anom}
\end{figure}

\begin{figure*}[t]
    \centering
    \includegraphics[width=0.9\linewidth]{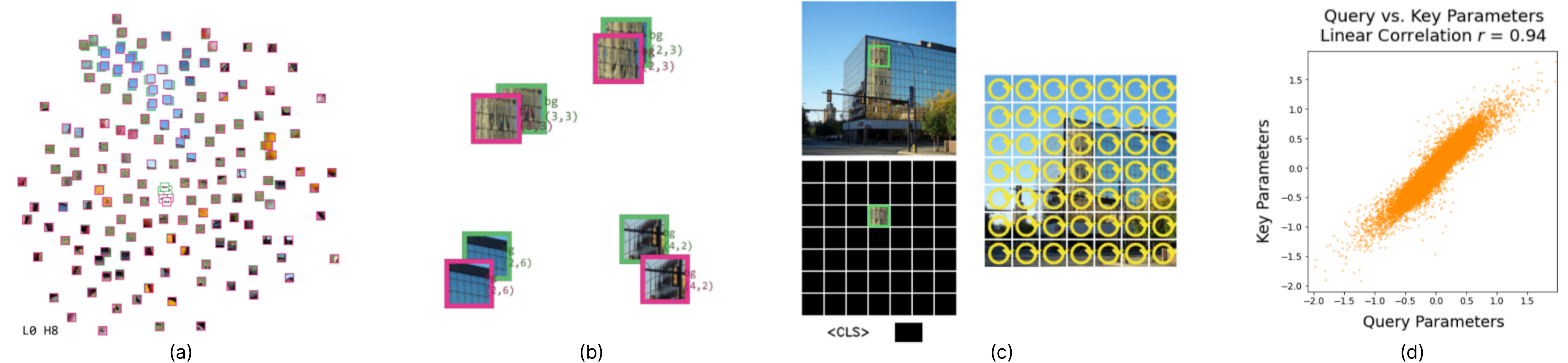}
    \caption{Identifying a ``look at self'' attention head in ViT-32. \textbf{(a)} Single View shows queries and keys are sparsely distributed in the joint embedding space. \textbf{(b)} Zooming in, query and key vectors of the same image token are tightly overlapped. \textbf{(c)} Image View reveals tokens pay most attention to themselves. \textbf{(d)} Comparing the learned parameters of query and key projection layers confirms that they learn redundant projections.}
    \label{fig:redundant-attention-head}
\end{figure*}

\textbf{Norm disparities and null attention.}
While exploring GPT-2 in Matrix View, 
we observed that 
in early model layers, some query and key clusters were well-separated, even after key translation (Sec.~\ref{sec:key_norm}). By coloring by norm (as measured before the norm scaling step), we saw that in many heads, there is a significant disparity between the norms of query and key vectors (Fig.~\ref{fig:gpt_anom}a). 
When query norms are small (\ie light green), key norms tend to be large (\ie dark pink), and vice versa. 
Computing the average norm difference between queries and keys in GPT-2 vs. BERT, we found that in the former, the mean query norm - key norm = \textbf{-4.59} across attention heads, while in the latter, the mean difference is only \textbf{0.41}.
None of our experts could explain this finding: ``It doesn't really make sense why queries and keys would have such different norms'' \textbf{(E6)}.
Interestingly, a paper published after we made this observation \cite{dehghani2023scaling} points to out-of-control query and key norms as a cause of serious training instability, indicating that this phenomenon may be worth studying further.
This observation inspired our scaling approach from Sec.~\ref{sec:key_norm} as well.

We also noticed that in many GPT-2 heads, most attention is directed to the first token (Fig.~\ref{fig:gpt_anom}b), especially in later layers. 
~\cite{vig2019analyzing} briefly mentions that the first token is treated as a null position for attention-receiving in GPT-2 ``when the linguistic property captured by the attention head doesn't appear in the input text.'' However, this phenomenon remains underexplored, proposing another open interpretability question to consider. \textbf{E2} and \textbf{E6} both noticed this anomalous behavior on their own with our tool, and all of our experts were surprised by this finding.
~\cite{voita2019analyzing} shows that pruning the majority of attention heads in transformers may not significantly impact model performance, which perhaps can be partially attributed to this dominant null attention pattern. Regardless, AttentionViz allows users to filter out attention paid to the first token, uncovering hidden query-key interactions.

\textbf{``Look at self'' attention heads.} AttentionViz can also reveal surprising attention patterns in vision transformers. In Matrix View, we identified several heads in early layers of ViT-32 with very diffused key-query clusters (Fig.~\ref{fig:redundant-attention-head}a). Looking at one such attention head (layer 0 head 8), we discovered that the query and key embeddings of the same token form a small but dense cluster, with each query-key pair well-separated from the others (Fig.~\ref{fig:redundant-attention-head}b). From the transparency heatmap in Image View, we see that the patch is solely attending to itself (Fig.~\ref{fig:redundant-attention-head}c). Switching to the arrowed attention lines, we discover that the overall attention pattern for this image is ``look at self,'' where no information is flowing between image tokens in this head.

After identifying this irregular attention pattern, we checked the learned parameters of the query and key matrices with a correlation test. 
We found a strong similarity score (linear correlation $= \textbf{0.94}$), indicating that the query and key layers in this ViT head are indeed learning redundant projections (Fig.~\ref{fig:redundant-attention-head}d). \textbf{E3} noted that this knowledge could 
inform model pruning experiments, and AttentionViz could similarly be used to detect potential training failures or other irregularities.

\subsection{Takeaways from User Feedback}\label{sec:feedback}
\textbf{Merits of Matrix View.} Several experts found the ``global'' perspective provided by Matrix View to be the most novel and valuable part of AttentionViz. As \textbf{E6} said, ``It's great for quick comparison and frees you from tuning hyperparameters when you want to visualize multiple embeddings at once.'' \textbf{E7} also mentioned that Matrix View is useful because ``for smaller visualizations, I can just code up something myself, but it's a lot harder at scale and with more data.'' These comments suggest that this idea of visualizing and comparing embeddings at scale may be beneficial in other ML settings as well. 

\textbf{Applications for joint query-key embeddings.} Experts proposed various use cases and extensions for our visualization technique, evidencing its wider applicability. 
For example, \textbf{E2} suggested visualizing patterns in untrained or corrupt transformers, and both \textbf{E3} and \textbf{E7} wanted to visualize changes in attention during training for their own models, aligning with our original goals (Sec.~\ref{sec:goals_tasks}). \textbf{E2} and \textbf{E6} also suggested adapting our tool to help with causal tracing, explaining that ``it might be useful to track attention flow throughout the model for hypothesis testing.'' Similarly, \textbf{E3} expressed interest in looking into ``how two attention patterns connect in different heads,'' which could certainly be applied to visualizing induction head pairs. \textbf{E2} noted that adding a way to ``quantify similarity between two heads'' could be useful, while \textbf{E6} proposed ``measuring or visualizing randomness in heads'' for model pruning purposes. 

\textbf{Embedding projections -- to trust or not to trust?}
\textbf{E3} highlighted the challenges of using projection methods. 
While they appreciated the striking geometric patterns (\eg spirals) we found, \textbf{E3} expressed some skepticism about interpreting these visualizations due to the distortion from techniques such as t-SNE and UMAP: ``How do I know if I can trust what I see?''
This emphasizes the importance of tying visual insights to actionable interventions, perhaps through augmenting our tool to support hypothesis testing in addition to exploration.

\textbf{Flexibility-usability tradeoff.}
\textbf{E2} indicated that AttentionViz ``feels very usable and customizable,''
contrasting with ``existing visualization tools that are too overwhelming to learn and use.''
However, some experts like \textbf{E6} were still worried that ``showing all the features and heads might be overwhelming... Is there a way to summarize the information? Or focus more on a specific task?'' 
\textbf{E7} added, ``I wonder if there's a quicker, more digestible way to label heads,'' suggesting an approach closer to feature visualization~\cite{olah2017feature}. 
We designed AttentionViz to be a flexible tool (\eg allowing attention analysis in different transformers and at different granularities), but it seems that the flexibility-usability tradeoff~\cite{lidwell2010universal} of our design could still be improved.

\textbf{Additional interaction modes.} 
Some experts suggested additional interaction modes, \eg on-the-fly inference \textbf{(E3)} or further dimensionality reductions on circled clusters of queries and keys to reveal additional information and perform fine-grained analyses \textbf{(E2)}. \textbf{E7} stressed the importance of allowing users to directly upload new datasets to the system: ``The tool could be even more powerful... people are going to want to explore more with it like adding their own images.''
\section{Conclusions \& Future Work}
In this work, we introduce a new technique for visualizing transformer self-attention based on a joint embedding space for queries and keys.  
Applying our technique, we create AttentionViz 
(demo: \url{http://attentionviz.com}), an interactive visualization tool, and use it to gain insights about attention in both language and vision transformers. For instance, we discover novel hue/frequency behavior in ViT, and striking query-key norm disparities in GPT-2.
Although our approach is tailored to self-attention, it can be generalized to other attention mechanisms (\eg cross attention) with proper modifications to vector normalization. Similarly, we focus on semantic patterns, but AttentionViz could be used to study syntactic features as well, \eg by using NLP libraries to add metadata such as part of speech.

Expert feedback also points to several avenues for future work, such as 
finding ways to manage the complexity of multiple embedding visualizations and focus users on features of interest. 
Our system is currently limited by data pre-computation times and memory requirements, but we plan to improve the scalability of AttentionViz and allow users to add new inputs on the fly. A sufficiently large random sample (\eg a few thousand tokens) is enough to surface large-scale, global phenomena, but larger datasets and models may reveal additional semantic insights, as suggested by~\cite{oquab2023dinov2}. Trying other data sampling approaches might be fruitful as well (\eg to reduce biases from Zipf's law \cite{zipf1932selected}).

Another natural direction for future research is exploring how to incorporate information from value vectors in each attention head~\cite{vaswani2017attention}. These value vectors are an essential part of the attention mechanism, though it is not clear how to visualize them in the context of queries and keys. Finding the right visualization approach might shed more light on how attention heads function. Finally, although AttentionViz is an exploratory tool, adapting it for hypothesis testing and/or causal tracing might provide support for practical model debugging.

\acknowledgments{%
We would like to thank all the participants in our user interviews for their time and invaluable insights. We also thank the anonymous reviewers for helping us improve this paper with their thorough and constructive comments. Finally, we are grateful for the thoughtful feedback and support provided by the members of the Harvard Insight + Interaction Lab throughout this project. %
}

\bibliographystyle{abbrv-doi-hyperref}

\bibliography{main}


\appendix 

\end{document}